\documentclass[doublecol,figures]{epl2}

\usepackage{amssymb}

\title{Quantum charge diffusion in underdamped Josephson junctions and
superconducting nanowires}
\shorttitle{Quantum charge diffusion in Josephson junctions and nanowires}

\author{A. Zazunov \and N. Didier \and F. W. J. Hekking}
\shortauthor{A. Zazunov \etal}

\institute{Universit\'e Joseph Fourier, Laboratoire de Physique et de Mod\'elisation des
Milieux Condens\'es, CNRS\\BP~166, 38042~Grenoble, France, EU}


\abstract{
The effect of quantum fluctuations on the current-voltage characteristics of Josephson
junctions and superconducting nanowires is studied in the underdamped limit. Quantum
fluctuations induce transitions between a Coulomb--blockade and a supercurrent branch,
and can significantly modify the shape of current-voltage characteristics in the case of
a highly resistive environment. Owing to the phase-charge duality, our results can be
directly extended to the opposite overdamped limit.
}

\begin{document}

\maketitle

%%%%%%%%%%%%%%%%%%%%%%%%%%%%%%%%%%%%%%%%%%%%%%%%%%%%%%%%%%%%%%%%%%%%%%%%%%%%%%%%%%%%%%%%%%%%%%%%%%%%%%%%%%

\section{Introduction}

A small-capacitance Josephson junction is a quantum system with rich dynamics.
The two conjugate variables are the superconducting
phase difference~$\phi$ across the junction and the charge~$Q$ on its electrodes.
Correspondingly, at low temperatures the behavior
of the junction is determined by the competition between the
Josephson energy~$E_J$ and the charging energy
$E_C = (2e)^2/ 2C$, where~$C$ is the junction capacitance~\cite{tinkham}.
If $E_J \gg E_C$, $\phi$ is well-defined and a phase-coherent
Cooper-pair current can flow through the junction in the absence
of an external voltage~$V$. In the opposite limit $E_J \ll E_C$ an
insulating state with a well-defined charge~$Q$ on the electrodes is possible.
At the same time, the dynamics of~$\phi$ and~$Q$
is crucially influenced by dissipation caused by the electromagnetic environment surrounding the junction.
Because of the mutual interplay of quantum mechanics, nonlinearity and dissipation,
the consistent theoretical description of Josephson junctions still remains far from being complete.

The dc current-voltage ($I$-$V$) characteristics of a Josephson junction embedded in a circuit of
resistance~$R$ have been well studied in the so-called
overdamped~\cite{tinkham} case corresponding to
small values of $R/R_Q$ ($R_Q =h/4e^2$ is the resistance quantum)
and the ratio $E_J/E_C$~\cite{IvanZilb,SchonZaikin,IngoldNazarov,zwerger,
IngoldGrabertEberhardt,GrabertIngoldPaul,IngoldGrabert}.
For small $R<R_Q$, the supercurrent peak at zero voltage acquires a finite width.
With increasing $R$, quantum fluctuations of the phase become more
important and the supercurrent peak gradually moves to higher
voltages. This corresponds to the transition (driven by the
environment) from superconducting behavior found for small $R$ to
a complete Coulomb blockade when $R> R_Q$.
Meanwhile, the opposite underdamped regime, which has been extremely
difficult to achieve experimentally, has attracted less attention.
However, recently experiments were performed~\cite{haviland,corlevi}
on junctions with $E_J/E_C > 1$, embedded in a tunable, highly resistive
environment, $R \gg R_Q$,
enabling the study of the same junction in different environments.
In particular, a voltage peak near zero current
followed by a back-bending to lower voltages at higher currents
was observed.
This is the so-called Bloch nose~\cite{likharev} which,
in accordance with a duality property~\cite{schmid,guinea,SchonZaikin,weiss},
resembles the $I$-$V$ characteristic of an overdamped junction but
with the role of voltage and current interchanged.
A quantitative comparison between theory and experiment
has been made in the classical limit where thermal fluctuations dominate~\cite{corlevi,bhp}.

In this Letter we study for the first time the influence of quantum fluctuations on the
$I$-$V$ characteristics of an underdamped Josephson junction. We show that the
quantum-mechanical nature of the electromagnetic environment can strongly modify the
crossover from Coulomb blockade to superconducting behavior. Without fluctuations the
$I$-$V$ characteristic contains a sharp cusp that connects two distinct branches: a
zero--current finite--voltage branch corresponding to Coulomb blockade and a supercurrent
branch corresponding to Bloch oscillations of the voltage. Thermal fluctuations induce
transitions between these branches thereby smearing the cusp. This thermal smearing can
be described using a Fokker-Planck approach~\cite{bhp}. Quantum effects show up at
temperatures comparable with the characteristic cutoff frequency $\omega_c$ of the
environmental modes. As we will detail below, at intermediate temperatures $k_BT \lesssim
\hbar\omega_c$, a quasiclassical regime exists. In this regime small quantum corrections to
the thermal smearing are found, described by the so-called quantum Smoluchowski equation.
In the limit of low temperatures $k_BT \ll \hbar\omega_c$, quantum fluctuations not only
smear the cusp but also shift its position. Specifically, with increasing dissipation
strength the voltage peak shifts to nonzero current:
quantum fluctuations induce tunneling events of flux quanta that disrupt the Bloch oscillations.

Due to the duality
symmetry, our results can be directly mapped to the case of an overdamped junction, where
we have found that the role of quantum fluctuations has not been adequately studied, for
instance, in the quantum Smoluchowski regime [see eq.~(\ref{QSE}) and discussion below].
Finally, we demonstrate that all our results also apply to the case of the
recently proposed quantum phase-slip junctions which are realized in superconducting
nanowires~\cite{MooijNazarov}.
%
%%%%%%%%%%%%%%%%%%%%%%%%%%%%%%%%%%%%%%%%%%%%%%%%%%%%%%%%%%%
\begin{figure}[t!]
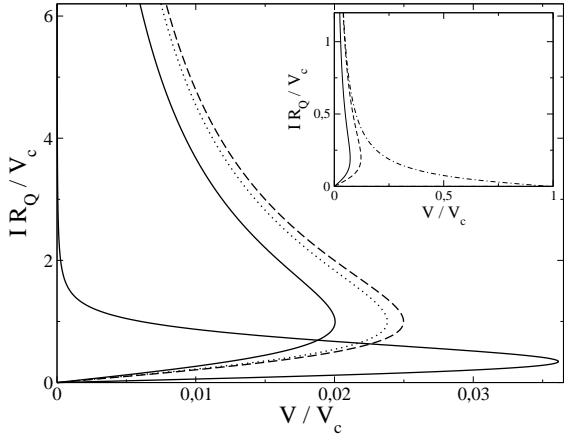

\onefigure[height=0.24\textheight]{fig1.eps}
\caption{
$I$-$V$ characteristics for $g$=0.1 and $\beta U_0$=0.1.
The dashed line corresponds to the classical expression~(\ref{classIV}) while the dotted line gives the quasiclassical solution from~(\ref{QSE}) for $\beta\hbar\omega_c$=1.
Solid lines depict the result from~(\ref{VfromCF})
with $\beta\hbar\omega_c$=10 for the rounded $I$-$V$
characteristic and $\beta\hbar\omega_c$=0.1 for the sharp one.
Inset: (from right to left) the classical limit without
fluctuations~(\ref{clasresult}), the result~(\ref{classIV}), and
the solution from~(\ref{VfromCF}) for $g$=0.1, $\beta U_0$=0.5,
and $\beta\hbar\omega_c$=50. } \label{figQS}
\end{figure}
%%%%%%%%%%%%%%%%%%%%%%%%%%%%%%%%%%%%%%%%%%%%%%%%%%%%%%%%%%%

%%%%%%%%%%%%%%%%%%%%%%%%%%%%%%%%%%%%%%%%%%%%%%%%%%%%%%%%%%%
\begin{figure}[t!]
\onefigure[height=0.24\textheight]{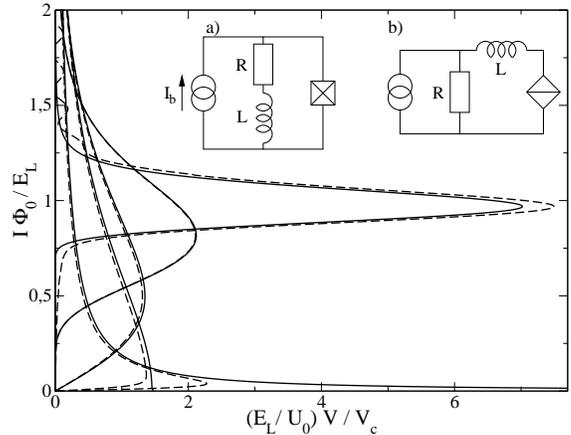}
\caption{
$I$-$V$ characteristics at $T=0$ for $\hbar \omega_c = 100 \, U_0$. From top to bottom, solid lines
represent~(\ref{PE}) with $g$=100, 5, 1, 0.5 and 0.1. Dashed lines
are the corresponding curves for the solution from~(\ref{VfromCF}) with $\beta U_0$=0.5 and $\beta\hbar\omega_c$=50.
While decreasing~$R$ the Bloch nose is shifted to the
finite current $I=\Phi_0/2L$, where
the tunneling of a flux quantum
requires an energy exchange with the bath.
Insets: equivalent circuits of current-biased Josephson junction~(a)
and nanowire-based QPS~(b) with the diamond symbol representing the phase-slip process.
}
\label{figPE}
\end{figure}
%%%%%%%%%%%%%%%%%%%%%%%%%%%%%%%%%%%%%%%%%%%%%%%%%%%%%%%%%%%

\section{Model}

We start our analysis by considering a current-biased Josephson junction (see inset a of fig.~\ref{figPE}).
The junction is shunted by a resistance~$R$ and biased by an external dc
current~$I_b$. As long as quasiparticle excitations are neglected,
the system is described by the following Hamiltonian:
\begin{equation}
H = \frac{\left( Q - Q_x \right)^2}{2C} - E_J \cos \phi - \frac{\hbar}{2e} I_b \, \phi + H_\mathrm{bath} \, ,
\end{equation}
where the operators  $Q$ and $\phi$ obey the commutation relation $\left[ \phi , Q \right] = 2 e \mathrm{i}$.
The fluctuating charge~$Q_x$ is associated with the current $I_x =
\dot{Q}_x$ flowing through the shunting resistor. The latter is modeled by a bath of
harmonic oscillators with frequencies $\left\{ \omega_\alpha \right\}$, which is
described by $H_\mathrm{bath} = \sum_\alpha \hbar \omega_\alpha \left( p_\alpha^2 +
x_\alpha^2 \right)/2$. In this model, $Q_x$ is represented by a weighted sum of
oscillator coordinates~$\left\{ x_\alpha \right\}$, $Q_x = 2 e \sum_\alpha \lambda_\alpha
x_\alpha$, while the influence of the bath on the junction dynamics is entirely
determined by the weighted spectral function, $K(\omega) = (\pi/2) \sum_\alpha
\lambda_\alpha^2 \delta(| \omega | - \omega_\alpha)$. As follows from the equations of
motion generated by~$H$, the choice $K(\omega) = R_Q \mathrm{Re} Y(\omega) / 2 \pi \omega$
with $Y(0) = 1/R$ reproduces the linear response (Ohm's law) of current~$I_x$ to the
voltage drop $V = (\hbar / 2e) \dot{\phi}$ on the resistor, $I_x(\omega) = Y(\omega)
V(\omega)$ (for Fourier transforms). In the case of interest here, $E_J \gg E_C$, it is
advantageous to switch to the Bloch-band description of the Josephson
junction~\cite{likharev}. Assuming that the junction dynamics is confined to the lowest
energy band $\epsilon_0(q) = - U_0 \cos (\pi q/e)$, we arrive at the tight-binding (TB)
model of our system with the Hamiltonian:
\begin{equation}
H_\mathrm{TB} = - U_0 \cos (\pi q/e - \chi) - (\hbar/2e)I_b \phi + H_\mathrm{bath} \, ,
\label{HamTB}
\end{equation}
where $q$ is the operator corresponding to the quasicharge
and $U_0=4\sqrt{2/\pi}E_C\left(2 E_J/E_C\right)^{3/4}\exp\left(-4\sqrt{2E_J/E_C}\right)$
is the Bloch bandwidth. The reduced fluctuating charge
\begin{equation}
\chi = 2 \pi \sum_\alpha \lambda_\alpha x_\alpha
\label{chi}
\end{equation}
gives rise to the bath correlation function
$J(\tau) = \langle \left( \chi(\tau) - \chi(0) \right) \chi(0) \rangle$, or explicitly,
\begin{equation}
J(\tau) = 2 R_Q \int_{- \infty}^{+ \infty} \upd \omega \,
\frac{\mathrm{Re} Y(\omega)}{\omega}  \,
\frac{\mathrm{e}^{- \mathrm{i} \omega \tau} - 1}{1 - \mathrm{e}^{-\beta \hbar \omega}} \, ,
\label{Jtau}
\end{equation}
where $\beta =1/k_B T$ is the inverse temperature.

Hamiltonian~(\ref{HamTB}) describes the quasicharge dynamics
in the lowest Bloch band under the influence of the
current $I = I_b - I_x$. The group velocity associated with the quasicharge,
$\upd\epsilon_0/\upd q = (\pi U_0/e) \sin (\pi q/e)$,
corresponds to the voltage across the junction.
Hence for a resistive environment Ohm's law yields $I_x  = (V_c/R) \sin (\pi
q/e)$, where $V_c = \pi U_0/e$ is the maximal (critical) voltage the junction can sustain.
Classically, the problem is equivalent to the equation of motion~\cite{likharev},
\begin{equation}
\dot{q} = I_b - (V_c/R) \sin (\pi q/e) \, ,
\label{classeq}
\end{equation}
describing overdamped quasicharge diffusion with damping rate
$\pi V_c/eR$. A stationary solution $\dot{q} = 0$ exists if $I_b <
V_c/R$: all the current flows through the resistor, while the
junction stays in a zero-current Coulomb blockade state with a
voltage drop $V=R I_b$. If $I_b > V_c/R$, $\dot{q} \ne 0$ and a
dynamical state exists at finite current with Bloch oscillations
of the voltage. From eq.~(\ref{classeq}) one finds the frequency of
the oscillations to be
$f_B = I/2e$;
by direct integration of
eq.~(\ref{classeq}) over one period one obtains the dc voltage~\cite{likharev}
\begin{equation}
V = R I_b-\sqrt{\left(R I_b \right)^2-V_c^2} \, .
\label{clasresult}
\end{equation}
The resulting $I$-$V$ characteristic corresponds to the aforementioned Bloch nose and is
depicted in fig.~\ref{figQS}, inset.

We now turn to the effect of fluctuations and introduce a cutoff of the bath spectrum at
frequency~$\omega_c$, chosen to be smaller than the gap 
$\omega_g = \sqrt{2 E_J E_C}/\hbar$
between the lowest Bloch bands
but higher than the Bloch bandwidth.
Furthermore, the competition between the cutoff energy and the temperature determines the nature, rather classical or quantum, of the environment.
Indeed, starting from a classical thermal bath when $k_BT\sim\hbar\omega_c$,
quantum fluctuations appear in the quasiclassical region $k_BT\lesssim\hbar\omega_c$
and become dominant at low temperatures $k_BT\ll\hbar\omega_c$.
Assuming that the effective impedance seen by the
junction is given by a resistance~$R$ in series with an
inductance~$L$, we can set $Y(\omega) = 1/(R -
\mathrm{i} \omega L)$, leading to $\omega_c = R/L$. Then, from
evaluating the integral (\ref{Jtau}) one obtains:
\begin{equation}
J(\tau) = - \mathrm{i} \, \mathrm{sign}(\tau) \, A(\tau) - M(\tau) \, ,
\end{equation}
with
\begin{eqnarray}
A(\tau) &=& \pi g \left( 1 - \mathrm{e}^{-\omega_c |\tau|} \right) \, ,
\label{A}
\end{eqnarray}
\vspace{-5mm}
\begin{eqnarray}
M(\tau) &=& 2 g \left[ \frac{\pi |\tau|}{\beta \hbar} -
\frac{\pi}{2} \cot\!\left( \frac{\beta \hbar \omega_c}{2} \right)
\left\{ 1 - \mathrm{e}^{-\omega_c |\tau|} \right\} \right.
\nonumber \\
&&\phantom{2 g \left[\right. }\left.+ \,
\sum_{k=1}^{+\infty} \frac{1 - \mathrm{e}^{-\nu_k |\tau|} }{k \left( 1 - \nu_k^2 / \omega_c^2 \right)}
\right] \, ,
\label{M}
\end{eqnarray}
where $g = R_Q / R$ is the dimensionless conductance, and $\nu_k = 2 \pi k / \hbar \beta$
is a Matsubara frequency.

Before proceeding, we want to comment on 
the magnitude of the effective inductance $L$ which has been associated with the cutoff $\omega_c$.
The restriction $\omega_c < \omega_g$ imposes a lower boundary on the values of $L$
to be consistent with our assumption of single--band charge dynamics:
\begin{equation}
L > \frac{2 \pi g^{-1}}{\sqrt{2 E_C / E_J}} \, L_J \, ,
\label{Lcondition}
\end{equation}
where $L_J = \left( \Phi_0 / 2 \pi \right)^2 / E_J$ is the Josephson inductance and
$\Phi_0=h/2e$ is the flux quantum. For relatively large $g > 1$, 
the condition (\ref{Lcondition}) can be satisfied even 
for an autonomous Josephson junction, where $L \sim L_J$ corresponds to the Bloch inductance~\cite{zorin}.
In the opposite limit of small $g \ll 1$, an experimental realization of this model would be
the use of an environment composed of Josephson junction SQUID arrays, whose effective
inductance can be tuned to relatively large values by magnetic flux~\cite{agren}.
Another option includes the use of a quantum phase--slip junction where
there is no limitation on $\omega_c$, as discussed at the end of the paper.

\section{Current-voltage characteristics}

To determine the $I$-$V$ characteristics, we calculate the average
of the operator for the voltage across the junction
\begin{equation}
V(t) = V_c \, \langle \, \mathcal{U}_C(t,0) \, \sin \theta(t) \rangle \, ,
\end{equation}
where
$\theta = (\pi / e) \, q - \chi$. The time evolution operator (in the interaction picture),
\begin{equation}
\mathcal{U}_C(t,0) = T_C \exp \left\{ ( \mathrm{i} U_0 / \hbar) \int_0^t \!\upd \tau \, \sum_{s = \pm} s
\, \cos \theta^s(\tau) \right\} \, ,
\end{equation}
is defined along the Keldysh
contour with the Keldysh index $s = +/-$ refering to the
forward/backward branch of the contour, and $T_C$ denotes the
Keldysh time-ordering. Expanding $\mathcal{U}_C$ in $U_0$, averaging
over the bath and performing summation over the Keldysh indices yield:
\begin{eqnarray}
\hspace{-13mm}&&
V(t) =
(V_c/2 \mathrm{i}) \sum_{n = 0}^{+ \infty} (-1)^n \left( U_0 / \hbar \right)^{2n + 1}
\nonumber \\
\hspace{-13mm}&&\phantom{V(t)}\times \int_0^t \!\upd\tau_1 \, \dots
\int_0^{\tau_{2n}}\!\upd\tau_{2n + 1} \sum_{\{ f_k \}}
\left( \prod_{k=1}^{2n + 1} \sin G_k \right) \mathrm{e}^{\Gamma_n} \, ,
\label{vafterbath}
\end{eqnarray}
where for given $n$ the discrete variables $f_k$ satisfy
\begin{equation}
|f_{k+1} - f_k| = 1
\end{equation}
with the constraint $f_0 = f_{2 n + 2}=0$ \cite{lenard, zwerger}, and
\begin{eqnarray}
G_k &=& A_{k, k-1} f_k + \sum_{k'=1}^{k-1} \delta A_{k k'} f_{k-k'} \, , \\
\Gamma_n &=& - \mathrm{i} (\pi / e) I_b \sum_{k=1}^{2n + 1} \left( \tau_k - \tau_{k-1}\right) f_k
\nonumber \\
&&-\sum_{k=1}^{2n + 1} M_{k, k-1} f_k^2  +
\sum_{k=2}^{2n + 1} \sum_{k'=1}^{k-1} \delta M_{k k'} f_k f_{k'} \, .
\end{eqnarray}
Here $\delta M_{k k'} = M_{k k'} + M_{k-1, k'-1} - M_{k-1, k'} - M_{k, k'-1}$,
$\delta A_{k k'} = A_{k, k-k'-1} - A_{k, k-k'}$, with the shorthand notation
$A_{k k'}=A(\tau_k-\tau_{k'})$, and similar for $M_{k k'}$.
The above equations are in a form
suitable to apply the ``nearest-neighbor approximation'' (NNA), where one assumes that
$\delta M_{k k'} \approx 0$ and $\delta A_{k k'} \approx 0$ \cite{zwerger}
on the relevant time scale $\tau > 1/\omega_c$.
Then, in the limit $t \rightarrow \infty$, the expression~(\ref{vafterbath}) can be evaluated exactly,
and one obtains for the dc voltage~$V$ across the junction:
\begin{equation}
V = V_c \, \mathrm{Im} \left( W_1 / W_0 \right)  \, ,
\label{VfromCF}
\end{equation}
where $W_n$ obey the following recurrent relation ($n \geq 1$)
\begin{equation}
\mathrm{i} a_n \left( W_{n - 1} - W_{n+1} \right) = W_n \, ,
\label{recurr}
\end{equation}
\begin{equation}
a_n = \frac{U_0}{\mathrm{i} \hbar} \int_0^{+\infty} \!\upd \tau \, \sin \left( n A(\tau)
\right) \, \mathrm{e}^{\mathrm{i} (\pi/e) I_b n \tau - n^2 M(\tau)} \, .
\label{an}
\end{equation}
Eqs.~(\ref{VfromCF})--(\ref{an}) constitute the central result of this Letter. The
quantities $W_n$ should be identified with the Fourier transform of the quasicharge
distribution function $W(q)$ in the steady state,
\begin{equation}
W(q) = \sum_{n = -\infty}^{+\infty} \, \mathrm{e}^{-\mathrm{i} n \pi q/e } \, W_n  \, ,
\end{equation}
with the property $W_{-n} = W_n^\ast$. This becomes
obvious by noticing that eq.~(\ref{VfromCF}) can be viewed as a result of averaging
the voltage operator with the quasicharge distribution~$W(q)$,
\begin{equation}
V/V_c = W_0^{-1} \int_{-e}^{+e} \!\upd q \, W(q) \sin \left( \pi q /e \right) \, .
\end{equation}

\section{Quasiclassical limit}

To shed light on the range of applicability of the NNA
for our system, we first study the quasiclassical limit where the
typical time scale of the quasicharge dynamics, determined by the
damping rate $g U_0 / \hbar$ and the frequency of Bloch
oscillations~$f_B$ [see eq.~(\ref{classeq})], is slow. Indeed, if
\begin{equation}
g U_0 / \hbar , \ f_B  \ll  1/\hbar \beta ,\ \omega_c \, ,
\label{Smoluchrange}
\end{equation}
one can neglect the terms $\mathcal{O}\!\left( \mathrm{e}^{- \nu_k
\tau},\,\mathrm{e}^{- \omega_c \tau} \right)$ in eqs.~(\ref{A})
and (\ref{M}). We thus approximate [$\Theta(\tau)$ is the unit
step function]
\begin{equation}
A(\tau) = \pi g \,\Theta(|\tau| - 1/\omega_c) \, ,
\label{Aapprox}
\end{equation}
and
\begin{equation}
M(\tau) = \left( g \nu_1 |\tau |  + \lambda \right) \Theta(|\tau| - 1/\omega_c) \, ,
\label{Mapprox}
\end{equation}
with
\begin{equation}
\lambda = 2 g \left[ \,
\gamma + \frac{\pi}{\beta \hbar \omega_c} + \psi\!\left( \frac{\beta \hbar \omega_c}{2 \pi} \right) \, \right] \, ,
\label{lambda}
\end{equation}
where $\gamma = 0.577...$ is the Euler constant and $\psi(x)$ is
the digamma function. Introducing the short--time cutoff for the
approximated $A(\tau)$ and $M(\tau)$ is the simplest way to
provide that they vanish at $\tau = 0$ [see the exact expressions
(\ref{A}) and (\ref{M})], which is necessary for the consistency
of the applied NNA. Assuming also $g \ll 1$, one obtains for~$a_n$
entering the recurrent relation~(\ref{recurr}),
\begin{equation}
a_n = \frac{z}{2 \mathrm{i}} \frac{\mathrm{e}^{-\Lambda n^2}}{n - \mathrm{i} \eta} \, ,
\label{anApprox}
\end{equation}
where $z = \beta U_0$, $\eta = (e/\pi) \beta R I_b$, and~\cite{eplQSE}
\begin{equation}
\Lambda = 2 g \left[ \,
\gamma + \frac{2 \pi}{\beta \hbar \omega_c} + \psi\!\left( \frac{\beta \hbar \omega_c}{2 \pi} \right) \, \right] \, .
\label{Lambda}
\end{equation}
Inspection of the terms dropped in the NNA reveals that in the range~(\ref{Smoluchrange}),
where $A(\tau)$ and $M(\tau)$ can be simplified by their asymptotic expressions,
the NNA becomes exact in the quasiclassical region where~$\lambda$ is small.
Note that for small $g$, the NNA can be applicable even at low temperature.

\section{Quantum Smoluchowski equation}

The crossover between the classical and quantum limit is
controlled by the parameter~$\lambda$ which is related to quantum corrections to the position dispersion
of a fictitious Brownian particle in the harmonic potential~\cite{prlQSE}.
The classical limit corresponds to $\beta \hbar \omega_c \sim 1$ and $g\ll1$, or equivalently $\lambda,\,\Lambda \rightarrow 0$.
The recurrences~(\ref{recurr}) can then be solved analytically in terms of the modified
Bessel functions, $W_n = I_{n - \mathrm{i} \eta}(z)$,
resulting in the $I$-$V$ characteristics obtained in ref.~\cite{bhp}:
\begin{equation}
V = R I_b - \frac{\sinh \left( \pi \eta \right)}{ e \beta \, |I_{\mathrm{i} \eta}(z)|^2} \, .
\label{classIV}
\end{equation}

Now expanding $a_n$ in eq.~(\ref{anApprox}) to the first order
in~$\Lambda$ to include quantum corrections,
one can derive from eq.~(\ref{recurr}) a differential equation for
the quasicharge distribution~$W(q)$,
\begin{equation}
\partial_q \left[ \hat{\mathcal{L}}(q, \partial_q) -
\Lambda \left( e / \pi \right)^4 U{'''}(q) \, \partial_q^2  \right] W(q) = 0 \, ,
\label{QSE}
\end{equation}
where $U(q) = - U_0 \cos (\pi q/e) - R I_b q$ is a washboard potential in the charge
variable, $U'(q) \equiv \partial_q U$.
The corresponding Smoluchowski differential operator,
\begin{equation}
\hat{\mathcal{L}}(q,\partial_q) = U{'}_{\hspace{-0.15cm} \mathrm{eff}}(q) + \beta^{-1} D(q) \, \partial_q  \, ,
\end{equation}
is renormalized by quantum fluctuations through both the effective potential
\begin{equation}
U_\mathrm{eff}(q) = U(q) + \Lambda (e/\pi)^2 \, U{''}(q)
\end{equation}
and the $q$-dependent diffusion coefficient
\begin{equation}
D(q) = 1 + 2 \beta \Lambda (e/\pi)^2 \, U{''}(q) \, .
\end{equation}
Eq.~(\ref{QSE}) is the so-called Quantum Smoluchowski Equation~(QSE)~\cite{prlQSE,eplQSE}
and constitutes the other main result of this Letter.
Eq.~(\ref{QSE}) describes the leading quantum corrections to the charge
dynamics which originate from the quantum nature of the bath. According to
eq.~(\ref{Lambda}), the influence of quantum fluctuations becomes more substantial with
increasing parameter $\beta \hbar \omega_c$.
The $I$-$V$ characteristics, parametrically dependent on the bias $I_b$ and calculated
for different $\beta \hbar \omega_c$, are shown in fig.~\ref{figQS}.
For $\beta \hbar \omega_c \gg 1$, quantum fluctuations reduce the
blockade voltage and facilitate a crossover to the Bloch oscillations.
Interestingly, for small $\beta \hbar \omega_c < 1$, the influence of thermal
fluctuations becomes suppressed by the inertia effect of the ``heavy'' Brownian particle
with an effective mass scaled as $\omega_c^{-1}$.
For a wide range of $I_b$, the junction is locked in the insulating state
because of the lack of energy exchange with the environment,
which results in a sharp crossover to the Bloch oscillations.
In other words, reducing the cutoff energy below the temperature
is equivalent to decoupling the junction from the environment (see fig.~\ref{figQS}).

We note in passing that the structure of the QSE obtained here
from the series expansion~(\ref{VfromCF}) is different from the one derived
in ref.~\cite{prlQSE} for a Brownian particle in a slightly anharmonic potential and later applied to
an overdamped junction~\cite{eplQSE}.
Therefore we cannot exploit the duality property (see below) to
treat the Smoluchowski range for underdamped junctions using the results of ref.~\cite{eplQSE}.
Another approximation consists of evaluating expression~(\ref{vafterbath}) using the asymptotes for long times~(\ref{Aapprox})
and~(\ref{Mapprox})~\cite{weiss}.
Such an approach leads to coefficients $a_n=\beta U_0\,\mathrm{e}^{-\lambda/2}\exp(-2\pi gn^2/\beta\hbar\omega_c)/2\mathrm{i}(n-\mathrm{i}\eta)$,
which tend towards the result~(\ref{anApprox}) in the quasiclassical region.
As a consequence, a QSE equivalent to the solution~(\ref{QSE}) is also recovered.

\section{Low temperatures}

We proceed by studying eqs.~(\ref{VfromCF})--(\ref{an})
in the low temperature limit, beyond the quasiclassical
region~(\ref{Smoluchrange}). At very low temperature, $\beta \hbar
\omega_c \gg 1$, the $I$-$V$ characteristics are entirely
determined by the first coefficient $a_1$, eq.~(\ref{an}),
consistent with the fact that the NNA is exact to the lowest order
in $U_0^2$. Closed-form analytical expressions can be obtained in
several cases. For instance, at finite temperatures and for small
values of $g$, the $I$-$V$ characteristic can be written as
\begin{equation}
V/V_c = u \, \frac{\left| \Gamma(g + \mathrm{i} \hbar \beta I_b/2e
) \right|^2}{\Gamma(2g)} \, \sinh \left( \pi \hbar \beta I_b/2e
\right) \,,
\end{equation}
where $u = \left( \beta U_0 / 4 \pi \right) \left( \beta \hbar
\omega_c \, \mathrm{e}^{\gamma} / 2 \pi \right)^{-2g}$ and
$\Gamma(x)$ is the Gamma function. The resulting linear resistance
$R$ varies as a power law with temperature, $R \propto
T^{2(g-1)}$, in agreement with the asymptotic analysis of
ref.~\cite{SchonZaikin}. At zero-temperature one recovers
\begin{equation}
V/V_c = (\pi U_0/2)\,P(hI_b/2e) \,, \label{PE}
\end{equation}
where $P(E) = 1/h \int \upd t \exp{[J(t) + iEt/\hbar]}$, dual to
the well-known result for the $I$-$V$ characteristic for incoherent
Cooper pair tunneling in an overdamped
junction~\cite{IngoldNazarov}. In fig.~\ref{figPE}, we plot
$I$-$V$ characteristics for different~$g$, both at zero and at
finite temperatures. As~$g$ increases, the voltage peak shifts to
finite values of the supercurrent. This behavior can be
interpreted in terms of incoherent tunneling of the
phase~\cite{averin90}. Indeed, for small values of~$g$, few
environmental modes are available. Consequently, only elastic
tunneling is allowed and a flat Bloch nose is recovered. When~$g$
is large, the equivalent circuit consists of a loop containing the
junction closed by the inductance~$L$. A phase-slip event occurs
when the energy to be released $\Phi_0 I_b$ corresponds to the
energy $\Phi_0^2/2L$ to add one flux quantum~$\Phi_0$ in the
loop, \textit{i.e.}, when $I_b=\Phi_0/2L$. At this finite current,
phase tunneling disrupts the Bloch oscillations and gives rise to
a voltage peak.

\section{Overdamped Josephson junctions}

A similar analysis within the Keldysh formalism can be achieved in the case of a voltage-biased
overdamped Josephson junction in series with a resistance $R$.
This circuit is conventionally described by the Hamiltonian
$H = \left( Q + Q_x \right)^2/2C - E_J \cos \phi - V_b Q_x + H_\mathrm{bath}$,
where $Q_x$ is a fluctuating charge on the junction capacitor and $V_b$ is the bias voltage.
Using an analysis of the equations of motion similar
to the one preceding eq.~(\ref{HamTB}) and performing canonical transformations,
one can show that the model description of the junction is equivalently given by
the weak-binding (WB) Hamiltonian
\begin{equation}
H_\mathrm{WB} = - E_J \cos \left( \phi - \chi/2 \pi \right) + V_b Q + H_\mathrm{bath} \, ,
\end{equation}
with $\chi$ the bath variable defined as before, eq.~(\ref{chi}).
In this representation, the junction capacitance~$C$ is encoded in terms of the bath parameters,
$C^{-1} = \left( R_Q/2\pi \right) \sum_\alpha \lambda_\alpha^2 \omega_\alpha$,
while the weighted spectral function of the bath is
given by $K_\mathrm{WB}(\omega) = 2 \pi \mathrm{Re} Z(\omega) / R_Q \, \omega$, with
$Z^{-1}(\omega) = 1/R - \mathrm{i} \omega C$.
The operator for the current flowing through the junction is given by
$I = I_c \sin \left( \phi - \chi/2 \pi \right)$,
where $I_c = 2 e E_J/\hbar$ is the critical current.
The Hamiltonians $H_\mathrm{TB}$ and $H_\mathrm{WB}$ are related by the following transformation:
\begin{equation}
\begin{array}{c}
\left( \, \pi q / e , \ \phi \, \right) \leftrightarrow \left( \, \phi , \ - \pi Q / e \, \right) \, ,\\[3mm]
U_0 \leftrightarrow E_J \, , \quad I_b \leftrightarrow V_b / R_Q \, , \quad \chi \leftrightarrow \chi / 2 \pi \, .
\end{array}
\end{equation}
Consequently, the series expansions for $V/V_c$ in the TB model and for $I/I_c$ in the WB model are dual:
we can transpose our results obtained for the quasicharge dynamics in
an underdamped junction onto the dual case of the phase dynamics in an overdamped junction.
A unified approach is thus provided, as well as new results such as the series expansion for
$I/I_c$ (see footnote~\footnote{
The WB expansion for $I/I_c$, which is dual to~(\ref{vafterbath}),
is similar but not identical to the one
obtained in ref.~\cite{GrabertIngoldPaul} using a different kind of approximation (not the NNA).}) 
and the quantum Smoluchowski equation.

\section{Superconducting nanowires}

We conclude by considering quantum phase-slip (QPS) dynamics in superconducting nanowires.
Based on a duality argument, ref.~\cite{MooijNazarov} suggests the following model Hamiltonian
to describe QPS events:
\begin{equation}
H_\mathrm{QPS}(\phi, Q) = E_L \left( \phi / 2 \pi \right)^2 - E_S \cos \left( \pi Q / e \right) \, ,
\end{equation}
where $E_S$ is an energy associated with the phase-slip process
which changes the phase difference $\phi$ over the nanowire by $2 \pi$,
and $E_L = \Phi_0^2/2L$ is an inductive energy of the wire with a kinetic inductance $L$.
Correspondingly, the Hamiltonian of a current-biased QPS junction
(see inset b of fig.~\ref{figPE}) can be written as
\begin{equation}
H = H_\mathrm{QPS}(\phi + \phi_x, Q) + \left( \hbar / 2e \right) I_b \, \phi_x +
H_\mathrm{bath} \, ,
\label{Hqps}
\end{equation}
where a fluctuating phase across the junction $\phi_x$ is related to
the voltage drop $V_R = R ( \dot{Q} - I_b )$ on the resistor,
$\dot{\phi}_x = (2e / \hbar) V_R$.
It is straightforward to see that the Hamiltonian (\ref{Hqps}) is exactly dual
to the Hamiltonian of a voltage--biased Josephson junction,
and can correspondingly be mapped onto the Hamiltonian~(\ref{HamTB})
with~$U_0$ replaced by~$E_S$, while the inductance~$L$ of nanowire is encoded in terms of the bath parameters,
$L^{-1} = \left( 2 \pi / R_Q \right) \sum_\alpha \lambda_\alpha^2 \omega_\alpha$.
With this device, the frequency $\omega_c=R/L$ results from the physical resistance and inductance of the wire,
providing a natural cutoff of the bath.
Our previous analysis for an underdamped Josephson junction,
and thus the $I$-$V$ characteristics, can be directly applied to
a superconducting nanowire (not being restricted by the TB limit).
A typical case of nanowire inductance $L \sim 1$ nH corresponds to
$\omega_c / 2 \pi \sim g^{-1} \times 1$~THz.
Assuming $E_S / h \gtrsim 10$~GHz \cite{MooijNazarov} and
$T \sim 1$~K, as follows from eq.~(\ref{Smoluchrange}), we estimate that the QSE range for nanowires
is relevant for $g \lesssim 0.1$. Note from the parameters used in fig.~\ref{figQS} that the quantum
fluctuations in QPS junctions should be substantial.

\section{Summary}

To summarize, we have studied the influence of quantum fluctuations on the $I$-$V$
characteristics of an underdamped Josephson junction. We have applied a unifying approach
based on the Keldysh formalism that enables us to obtain quantitative results for a wide
range of parameters. Using the NNA approximation, we show the significant role of quantum
fluctuations revealed both in the quasiclassical Smoluchowski regime and in the
low-temperature quantum regime. In the Smoluchowski regime, compared to the case of
thermal fluctuations, quantum fluctuations mainly lead to a renormalization of the
parameters describing the quasicharge dynamics of the junction~\cite{weiss}. The NNA
becomes exact in the case of incoherent phase-slip events at low temperatures. In this
limit, phase tunneling disrupts the Bloch oscillations, leading to a voltage peak at
finite current. The quantum effects are sensitive to both the dissipation strength and
cutoff frequency, and could be observed in experiments as in refs.~\cite{haviland,corlevi}
with a tunable environment. Besides Josephson junctions, our results are also relevant for
superconducting nanowires modeled as quantum phase-slip junctions at low temperature.

\acknowledgments

We thank B.~Dou\c{c}ot, H.~Grabert, W.~Guichard, L.~B.~Ioffe, P.~Schuck, and U.~Weiss for
useful discussions. Financial support from Institut universitaire de France and
IST-3-015708-IP EuroSQIP is gratefully acknowledged.

%%%%%%%%%%%%%%%%%%%%%%%%%%%%%%%%%%%%%%% BIBLIOGRAPHY %%%%%%%%%%%%%%%%%%%%%%%%%%%%%%%%%%%%%%%%%%%%%%%%%%

\end{document}